\documentclass[iop]{emulateapj}

\usepackage{graphicx}
\usepackage{multirow}
\usepackage{amsmath}
\usepackage{xcolor}
\usepackage{rotating}
\usepackage{threeparttable}

\shorttitle{HNC in Disks}
\shortauthors{Graninger et al.}

\begin{document}

\title{HNC in Protoplanetary Disks}

\author{Dawn Graninger, Karin I. \"{O}berg, Chunhua Qi}
\affil{Harvard-Smithsonian Center for Astrophysics, 60 Garden St, Cambridge, MA, 02138}
\email{dgraninger@cfa.harvard.edu}

\and
\author{Joel Kastner}
\affil{Center for Imaging Science, School of Physics \& Astronomy, and Laboratory for Multiwavelength Astrophysics, Rochester Institute of Technology, 54 Lomb Memorial Drive, Rochester NY 14623 USA}

\begin{abstract}
The distributions and abundances of small organics in protoplanetary disks are potentially powerful probes of disk physics and chemistry. HNC is a common probe of dense interstellar regions and the target of this study. We use the Submillimeter Array (SMA) to observe HNC 3--2 towards the protoplanetary disks around the T Tauri star TW Hya and the Herbig Ae star HD 163296. HNC is detected toward both disks,  constituting the first spatially resolved observations of HNC in disks. We also present SMA observations of HCN 3--2, and IRAM 30m observations of HCN and HNC 1--0 toward HD~163296. The disk-averaged HNC/HCN emission ratio is 0.1--0.2  toward both disks.  Toward TW Hya, the HNC emission is confined to a ring. The varying HNC abundance in the TW Hya disk demonstrates that HNC chemistry is strongly linked to the disk physical structure. In particular, the inner rim of the HNC ring can be explained by efficient destruction of HNC at elevated temperatures, similar to what is observed in the ISM. To realize the full potential of HNC as a disk tracer requires, however, a combination of high SNR spatially resolved observations of HNC and HCN, and disk specific HNC chemical modeling.
\end{abstract}


\section{Introduction}

The disks around young stars contain the gas and dust from which planets form. From theory \citep{Aikawa96,Bergin03} and observations \citep{Rosenfeld12a, Qi13}, disks are characterized by steep temperature and radiation gradients both radially and vertically.  These disk structures affect planet formation \citep{Takeuchi96, Kretke12}, motivating the quest for new emission-line diagnostics that can trace physical structure gradients in disks.

This study focuses on HNC, a common high density gas tracer in the ISM. At low temperatures, HNC forms efficiently together with HCN through ion-molecule reactions.  HNC and HCN are observed in a variety of different sources, but their relative abundances can vary by more than order of magnitude, which has been related to the physical environment where the emission originates. In pre-stellar and protostellar environments, the observed HNC/HCN abundance ratio is  $\sim$1 at 10 K and decreases as temperature increases (e.g.~\citet{Schilke92, Pratap97, Liszt01, Jorgensen04, Padovani11, Jin15arxiv}), suggesting that thermal reaction barriers play an important role in controlling this ratio. Work by \citet{Loison14} has shown that the HNC/HCN ratio also depends on the elemental abundances of carbon at low temperatures, due to destructive reactions with atomic carbon. In less shielded regions, the HNC/HCN ratio also depends on the radiation present \citep{Meijerink07}. 

Nearby, protoplanetary disks, such as the one surrounding TW Hya, provide an interesting test bed for evaluating the relative impact of different environmental parameters on the HNC chemistry, due to their relatively well characterized geometries and density, temperature and radiation structures. The HNC distribution in disks is interesting in its own right. HNC and HCN are two of very few  nitrogen-bearing molecules detected toward disks \citep{Dutrey97,Dutrey07, Chapillon12, Oberg15} and their distributions may therefore provide some of the best available constraints on the nitrogen reservoir. Cyanides and iso-cyanides are also important for some prebiotic chemistry scenarios \citep{Powner09}. Observing HNC in disks has proven challenging, however. The intrinsic disk scales are small and the molecular line fluxes are weak compared to interstellar clouds and protostars. To date, HNC has only been detected in a single disk, DM Tau, \citep{Dutrey97} and no spatially resolved observations exist.

In this letter, we report the first interferometric detection of HNC in two disks, around  the T Tauri star TW Hya (D = 54 pc \citet{Qi04}) and Herbig Ae star HD 163296 (D = 122 pc, \citet{Qi11}), using the SMA. We combine these observations with new and previous SMA observations of HCN toward TW Hya and HD 163296, and new IRAM 30m observations of HNC and HCN 1--0 toward HD 163296, with the aim to constrain the HNC/HCN chemistry in disks.

\section{Observations of HNC and HCN}

The HNC 3--2 line  at 271.981 GHz was observed toward the TW Hya disk ($\Delta\alpha$ = 11$^h$01$^m$51.875$^s$, $\Delta\delta$ = --34$^\circ$42$'$17.155$''$; J2000.0) on 2013 April 5 and 2014 April 8 with the SMA eight-antenna interferometer located on Mauna Kea, HI. Observations were taken using the compact (COM) and extended (EXT) arrays configurations (projected baselines of 6.3--164.8~m). The gain calibrator was J1037-295. Calisto and Titan were used as the flux calibrators and the bandpass response was calibrated using 3C279. The derived fluxes of J1037-295 were 0.79 Jy on 2013 April 5 and 0.70 Jy on 2014 April 8.

The same HNC line was observed toward the disk around HD 163296 ($\Delta\alpha$ = 17$^h$56$^m$21.28$^s$, $\Delta\delta$ = --21$^\circ$57$'$22.1$''$; J2000.0) on 2013 April 5 with the SMA in the compact (COM) array configuration with projected baselines of 7.7--62.1~m. J1733-130 was used as the gain calibrator. The bandpass response was calibrated using BLLAC. Flux calibration was done using observations of Neptune. The derived fluxes of J1733-130 was 1.34 Jy. 

The HCN 3--2 line at  265.886 GHz was observed toward HD 163296 on 2008 April 20 and May 12 with the SMA in the compact-north (COM-N) array configuration with projected baselines of 7.8--114.3 meter. J1733-130 was used as the gain calibrator. The bandpass response was calibrated using observations of 3C273. Flux calibration was done using observations of Titan and Callisto. The derived fluxes of J1733-130 were 2.50 Jy on 2008 April 20 and 2.13 Jy on 2008 May 12. 

All data were phase- and amplitude-calibrated using the MIR software package\footnote{http://www.cfa.harvard.edu/$\sim$cqi/mircook.html}. Continuum and spectral line maps were generated and CLEANed using the MIRIAD software package. Images were constructed from the TW Hya data using a Briggs (robust) weighting factor of 0 and in HD 163296 data using 2.0 and 0.3 for HCN and HNC, respectively. The resulting synthesized beam for HNC in TW Hya is $1\farcs8\times1\farcs0$. For HD 163296, the synthesized beams for HCN and HNC are $3\farcs0\times1\farcs9$ and $4\farcs5\times2\farcs5$, respectively.

HCN 1--0 and HNC 1--0 lines were observed as a part of a single-dish 3~mm line survey of the disk around HD 163296 on 2012 September 13 - 16 with the IRAM 30m telescope on Pico Veleta using the EMIR 90GHz receiver and the Fourier transform spectrometer (FTS) backend in average summer weather (pwv $\sim$ 4-8 mm). Two spectral settings were used with the four sidebands covering 84.0 to 115.8 GHz with a resolution of 200 kHz ($\sim$0.5 - 0.6 km/s). Pointing was checked every 1-2 hrs and focusing every 4 hrs. The data was acquired in the wobbler switching modes, with typical throws of 60-120$"$ and an on-source integration time of $\sim$12-15 hrs per spectral setting. Spectra were baseline subtracted and averaged using CLASS\footnote{http://www.iram.fr/IRAMFR/GILDAS}. To convert from antenna temperature, T$^*_a$, to main beam temperature, T$_{mb}$, forward and beam efficiencies of 0.95 and 0.81 were applied.

\section{Results}
\subsection{Observations}

\begin{figure}[htp]
\centering
\epsscale{1.1}
\plotone{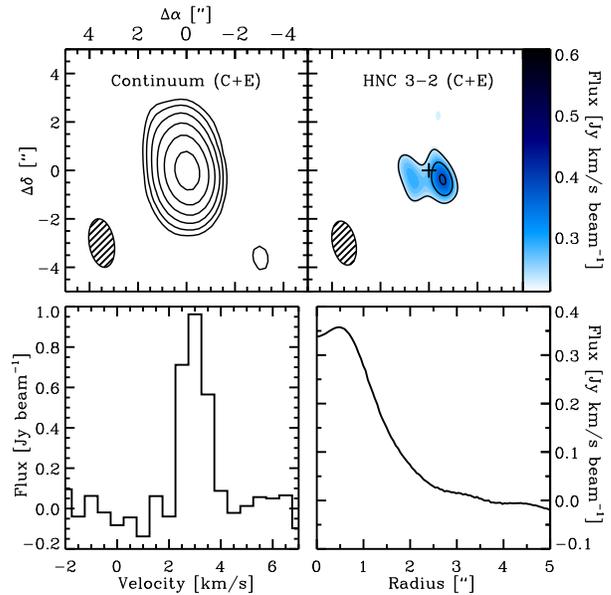}
\caption{The 1.1 mm continuum and HNC 3-2 emission (upper panels) from TW Hya using the compact + extended configurations of the SMA. Continuum contours are [10,20,40,...] mJy and the contours for HNC are [4,6,8]$\sigma$ with 1$\sigma$ = 0.064 Jy km s$^{-1}$ beam$^{-1}$. The lower left panel displays the spectrum obtained from the HNC 3-2 image. The lower right panel displays the deprojected, azimuthally averaged flux as a function of radius.\label{fig1}}
\end{figure}

HNC 3--2 emission is clearly detected and resolved toward TW Hya. Figure \ref{fig1} displays the 1.1 mm continuum and HNC 3--2 SMA observations, combining the compact and extended data. The resolved integrated continuum flux is 0.78 Jy which agrees with previous studies \citep{Qi13}. The velocity-integrated HNC flux peaks at 0.53 Jy km s$^{-1}$ beam$^{-1}$. The integrated HNC emission peaks are offset to the east and west of the continuum phase center, suggestive of a ring-like emission structure within the disk. In the HNC image the western flux peak is almost twice as strong compared to the eastern one, corresponding to $\sim3\sigma$. Asymmetries of this magnitude may be artifacts of CLEANing process and based on visual inspection alone we are agnostic about its significance. We evaluate this aspect of the the data when modeling the complex visibilities in \S3.2.

\begin{figure*}[htpb]
\centering
\epsscale{1}
\plotone{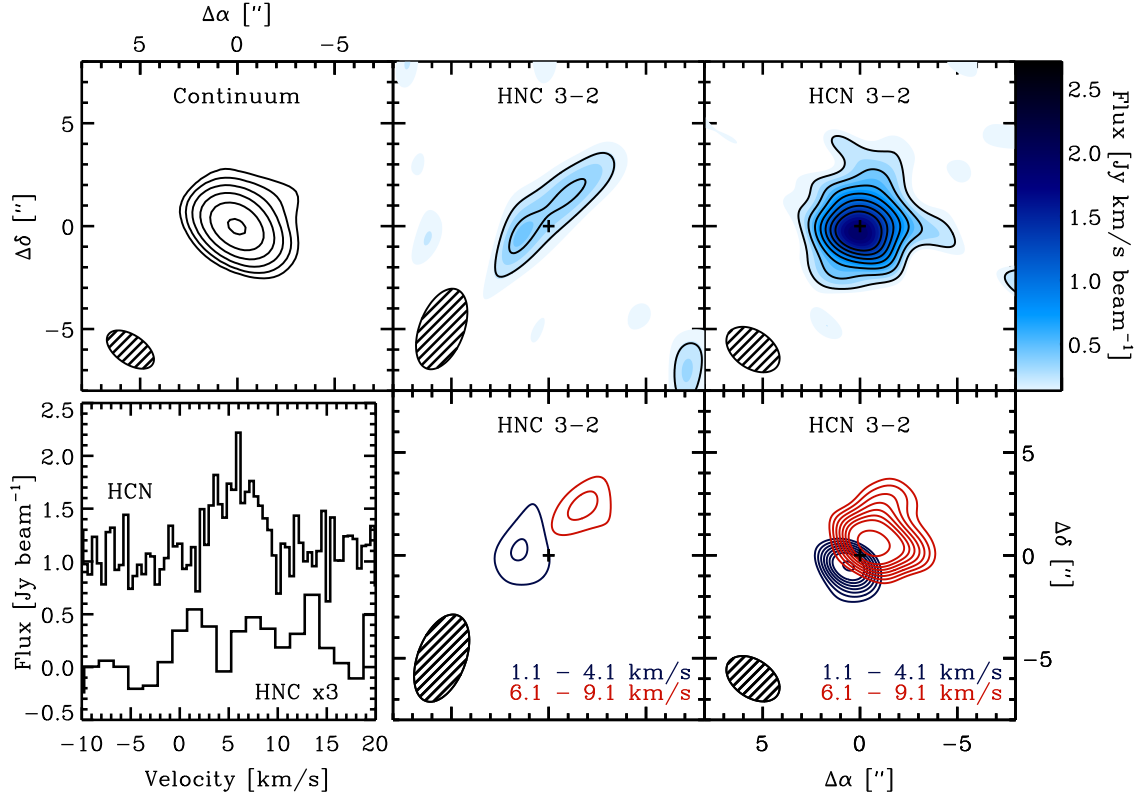}
\vspace{-3mm}
\caption{1.1 mm continuum (top left, contours=[20,40,80,...] mJy), integrated HNC 3--2 (top middle) and HCN 3--2 (top right) fluxes toward the disk around HD~162936 together with extracted spectra (bottom left) and integrated emission in two velocity bins (bottom middle and right). The HNC and HCN integrated flux contours are [2,3]$\sigma$ (HNC) and [3,5,7,..]$\sigma$ (HCN) with 1$\sigma=$0.21 and 0.14 Jy km s$^{-1}$ beam$^{-1}$, respectively. The HNC and HCN contours in the two velocity bins are [3,4]$\sigma$ (HNC) and [8,10,12,..]$\sigma$ (HCN).\label{fig2}}
\end{figure*} 

To isolate the radial flux variation in the HNC emission, Fig. \ref{fig1} (lower right panel) shows  the deporjected azimuthally averaged HNC line flux, using an inclination angle of 6$^{o}$ and position angle of 155$^o$ \citep{Qi04, Williams11}. The resulting emission profile reaches a maximum at $\sim$0.5$"$, indicating the presence of a small HNC hole at the center of the TW Hya disk. This agrees with the $\sim$1" separation between the two flux peaks in the HNC integrated emission image. The depth and size of the hole may both be affected by the image asymmetry, however. In \S3.2 we evaluate the evidence for an HNC ring and quantify its location based on an analysis in the ($u,v$)-plane.

The HNC spectrum in Fig. \ref{fig1} was extracted using a $2\farcs5\times2\farcs5$ box. The integrated line  flux is 1.2 Jy km s$^{-1}$ beam$^{-1}$ ($\sim$12$\sigma$). Using the the integrated HNC flux from this study and the HCN flux from \citet{Qi08}, the disk-averaged HNC/HCN  emission ratio is 0.18 $\pm$ 0.06, where the error is derived using the 1$\sigma$ rms for the two observations.

HNC is also detected toward HD 163296, but with a lower SNR. Figure 2 displays the 1.1 mm continuum, HNC, and HCN emission toward HD 163296 as well as the spectra of the molecules and the velocity structure of HNC and HCN. The velocity integrated emission maps yields a maximum HNC and HCN flux of 0.71 Jy km s$^{-1}$ beam$^{-1}$ and 2.71 Jy km s$^{-1}$ beam$^{-1}$, respectively. The HNC emission toward the HD 163296 disk is suggestive of a ring viewed at relatively high inclination, but the low SNR inhibits any further spatial analysis. 

Spectra were extracted using an elliptical mask, which increased the SNR compared to a square mask. The total integrated flux from these spectra is 0.66 Jy km/s beam$^{-1}$ ($\sim$4$\sigma$) and 4.47 Jy km/s beam$^{-1}$ ($\sim$9.5$\sigma$) for HNC and HCN, respectively. This yields an HNC/HCN ratio of 0.14 $\pm$ 0.05, similar to TW Hya.

\begin{figure}[h!]
\centering
\epsscale{1}
\plotone{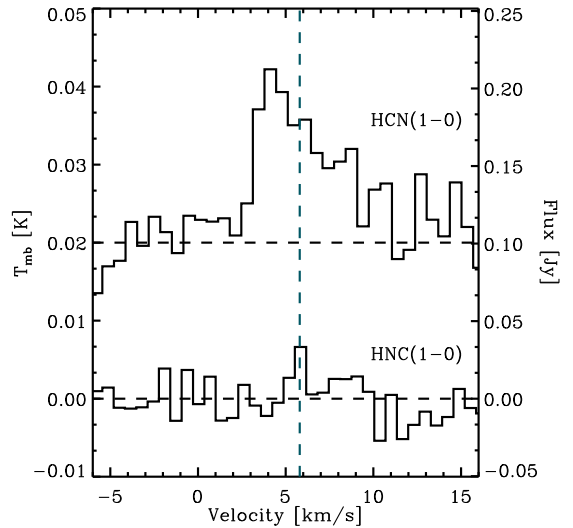}
\caption{IRAM 30m spectra of HNC and HCN 1--0 in HD 163296. The dashed line at 5.8 km/s is the systemic velocity of HD 163296.\label{fig3}}
\end{figure}

The IRAM 30m observations of HNC and HCN 1--0 toward HD 163296 should be more sensitive to the coldest disk material (Fig. \ref{fig3}). The integrated intensities of  HNC and HCN were found to be $<$0.033 K km s$^{-1}$ (3$\sigma$ upper limit) and 0.104 K km s$^{-1}$, respectively. This yields an upper limit on the HNC/HCN ratio of $<$0.3, consistent with the SMA data.

\subsection{HNC line modeling toward TW Hya}

\begin{figure*}[ht!]
\centering
\epsscale{1}
\plotone{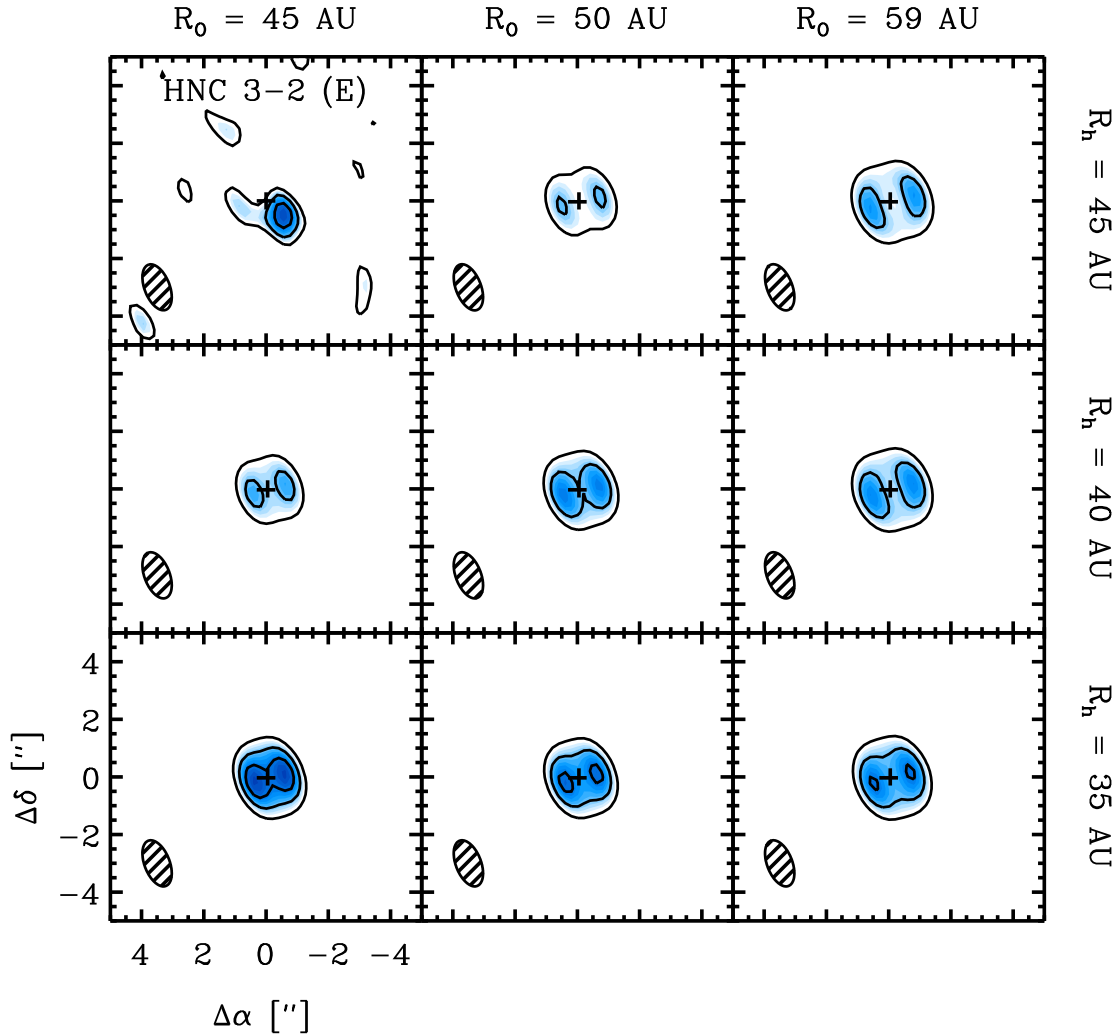}
\caption{Comparison of HNC 3--2 data (upper left panel) and models with an outer (R$_o$) and hole (R$_h$) radius. The contours are [3,4,5]$\sigma$, based on the observed RMS noise. Images are generated using the EXT data only to display the small-scale emission structures.\label{fig4}} 
\end{figure*}

To constrain the spatial extent of the HNC emission in TW Hya, we follow the procedure outlined in \citet{Qi08,Oberg12,Oberg15} to extract molecular column density profiles. First, parametric HNC abundance models are defined with respect to the TW Hya disk temperature and density structures from \citet{Qi13c}. The line emission is modeled using the two-dimensional non-LTE Monte Carlo radiative transfer code RATRAN \citep{Hogerheijde00} using collisional cross sections from \citet{Dumouchel10}. The HNC abundance models follow the form in \citet{Qi08}, which assumes a molecular layer with a vertically constant abundance, bound by surface ($\sigma_s$) and midplane ($\sigma_m$) boundaries defined with respect to the vertically integrated hydrogen column in the unit of 1.59$\times$10$^{21}$ cm$^{-2}$. We set the HNC boundaries using the values from \citet{Qi08} for HCN (log$\sigma_s$=-0.5, log$\sigma_m$=1.5). The radial distribution of HNC was modeled using an abundance power-law with an outer radius (R$_o$), hole radius (R$_h$), and power-law index of -1.5. The parameter R$_o$ was varied between 35 and 60 AU and R$_h$ was varied between 20 and 50 AU. The minimum $\chi^2$ value was calculated from the modeled and observed HNC emission in the ($u,v$)-plane using the data from the compact and extended configuration.

\begin{figure}[h!]
\centering
\epsscale{1}
\plotone{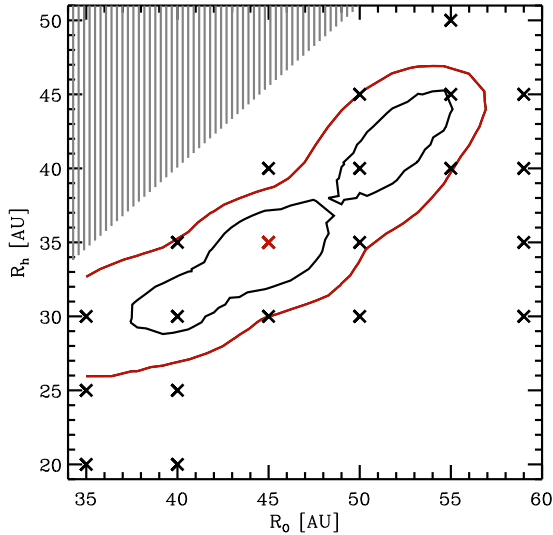}
\caption{$\chi^2$ contour plot for the best fit models. The contours are 2 and 3$\sigma$. The 3$\sigma$ contour is in red and the red cross marks the best-fit model. The black crosses mark the values of R$_h$ and R$_o$ modeled and the shaded region indicates where R$_h$$\geq$R$_o$.}
\label{fig5}
\end{figure}

Figure \ref{fig4} shows examples of emission models and observations in the extended configuration. The best fit model based on the $\chi^2$ value is a 10 AU wide ring with a R$_o$ of 45 AU and R$_h$ of 35 AU (Figure \ref{fig5}), shown in the lower left panel of Figure \ref{fig4}. This model results in no significant residuals when subtracted from the data in the ($u,v$)-plane and then imaged. The observed HNC azimuthal asymmetry toward TW Hya is thus not significant. The presence of a HNC ring in the TW Hya disk is by contrast a statistically significant result. Figure \ref{fig5} shows the models that fit the data within 2 and 3$\sigma$. Within 2$\sigma$, the hole size is  28--45~AU (3$\sigma$: 26--47~AU) and the outer edge is at 37--55~AU (3$\sigma$: $<$57~AU).

\section{Discussion}

In this study, we detect and spatially resolve HNC emission toward TW Hya and HD 163296. From these observations we extracted disk averaged HNC/HCN emission ratios, and the inner and outer HNC emission radii in the TW Hya disk. Here, we will discuss the constraints that these observations provide on the HNC chemistry and the disk structure.

\subsection{HNC/HCN Emission Ratios}

The disk-averaged HNC/HCN emission ratio is 0.1 -- 0.2 toward both TW Hya and HD 163296. Based on ISM observations \citep{Schilke92} and astrochemistry models \citep{Graninger14}, the HNC/HCN ratio can be used to constrain the characteristic temperature of the emission region because of  efficient destruction of HNC by hydrogen atoms at temperatures $>$25~K. However, to apply this result straightforwardly requires that HNC and HCN are co-spatial, that there is a simple relationship between emission and abundance ratios of HNC and HCN, and that temperature dominates the chemistry, none of which clearly hold in the case of these disks.

First, HNC and HCN are not co-spatial -- some HCN is present toward the center of the disk, where HNC is absent. This may be partially mitigated by averaging the HNC and HCN emission only across disk areas where HCN and HNC emission are respectively detected. Second, the relationship between line emission and abundance ratios is complicated by high HCN line opacities and different HCN and HNC excitation properties \citep{Sarrasin10}. We used the large velocity gradient radiative transfer code, RADEX \citep{vanderTak07}, with the collisional rate constants from \citet{Dumouchel10} to evaluate relationship  between the HNC/HCN line emission and abundance ratios using temperatures and densities characteristic of HCN disk emission \citep{Walsh10} and the HCN column density from \citet{Qi08}. We find that the line and abundance ratios can differ by up to an order of magnitude, demonstrating that more detailed modeling is required to extract HNC/HCN abundance ratios from the observed line emission.

Even if abundance ratios are successfully extracted, it may be difficult to infer a characteristic emission region temperature. Other environmental characteristics than temperature may regulate the HNC/HCN abundance ratio in disks. The model by \citet{Graninger14} neglects the influence of radiation, which is reasonable in cloud cores, but may not be a good approximation in disks \citep{Meijerink07}. \citet{Loison14} has also shown that the HNC/HCN ratio depends on the C/O ratio, which may be elevated in disks. In the absence of detailed disk chemistry models that incorporate the complete HNC/HCN chemistry, disk-averaged line ratios are thus not useful probes of disk properties. 

\subsection{The HNC inner and outer radii in the TW Hya disk}

In the disk toward TW Hya, there is a lack of HNC emission interior to 28--45~AU, corresponding to temperatures above 21--13~K using the \citet{Qi13c} temperature profile. HCN does not have an emission hole \citep{Qi08}. The observed absence of  HNC close to the star is consistent with temperature dependent HNC destruction \citep{Graninger14}. It is difficult to explain the hole with the increased UV flux present at smaller radii, since HNC and HCN are thought to have similar photodissociation cross-sections \citep{Roberge91, VanDishoeck06}.

HNC also has an outer emission radius between 35--55 AU, with the best fit at 45 AU, substantially smaller than the observed radii of other gases, including CO, HCO$^+$ and HCN \citep{Qi08}. This outer cutoff in HNC emission was unexpected. It coincides with the onset of C$_2$H emission, which is proposed to be a UV photochemistry product \citep{Kastner15}, and is close to the radius where the dust density sharply decreases \citep{Andrews12}.  These spatial coincidences suggest three potential explanations for the outer cutoff of HNC emission: 1. UV photodissociation of HNC due to deeper penetration of stellar and interstellar UV photons in the low-column density outer disk, 2. an increased UV photodissociation of H$_2$ and carbon-bearing molecules accelerating the HNC+H and HNC+C destruction processes at a given temperature \citep{Graninger14,Loison14}, and 3. grain-gas thermal decoupling due to lower gas and grain densities, causing a temperature reversal in the outer disk and thus the increase in thermally controlled HNC destruction (i.e. a similar process that sets the inner edge).

The first two explanations both depend on a more efficient UV photochemistry exterior to 45~AU, consistent with the onset of C$_2$H emission at the same radius \citep{Kastner15}, and signs of efficient UV photochemistry in the outer low-density regions of other disks \citep{Oberg11a}. Flared disks can intercept stellar UV radiation out to larger radii, and Ly$\alpha$ emission is efficiently down-scattered through disk towards the midplane \citep{Fogel11}. Additionally, the interstellar radiation field will contribute a non-negligible amount of UV radiation to the outer disk \citep{Bergin03}. Efficient UV penetration at large disk radii may increase the HNC photodissociation rate, reducing its abundance. This scenario could be tested by comparing HNC and HCN emission profiles. HNC and HCN are believed to have similar photodissociation cross-sections \citep{Roberge91, VanDishoeck06}, and if their outer emission radii are UV controlled, they should be the same. The best-fit outer HCN emission radius is $\sim$100~AU \citep{Qi08}, twice that of HNC, but this value is not well enough constrained to exclude direct photodissociation of HNC as an explanation for the HNC outer radius.

The second and third proposed explanations for the HNC outer cut-off should produce a smaller HNC than HCN emission radius, due to an increased HNC-to-HCN conversion rate in the outer disk. In the second scenario and increased UV penetration increases the atomic H and C abundance through H$_2$ and CO photodissociation \citep{Akimkin13}, speeding up HNC isomerization at a given temperature \citep{Graninger14}. In the third scenario, a lower density in the outer disk reduces gas-grain thermal coupling sufficiently that direct gas heating by the interstellar radiation field becomes important. This heating could result in a radial temperature gradient reversal outside of 45~AU, again increasing the HNC isomerization rate.

Higher spatial resolution observations with ALMA should be able to distinguish between the three scenarios. The relative sizes of the HNC and HCN outer radii should differentiate between \#1 on  one hand and \#2 and \#3 on the other. If \#1 is ruled out, the radial temperature profile of HCN  beyond 45~AU should be able to differentiate between \#2 and \#3. Whichever scenario is confirmed, spatially resolved HNC/HCN emission modeled using non-LTE techniques clearly has a large potential to probe interesting disk properties.

\acknowledgements
This work has benefited from discussions with Sean Andrews, David Wilner, Ryan Loomis, Eric Herbst and the helpful comments from the anonymous referee. The Submillimeter Array is a joint project between the Smithsonian Astrophysical Observatory and the Academia Sinica Institute of Astronomy and Astrophysics and is funded by the Smithsonian Institution and the Academia Sinica. The authors wish to recognize and acknowledge the very significant cultural role and reverence that the summit of Mauna Kea has always had within the indigenous Hawaiian community.  We are most fortunate to have the opportunity to conduct observations from this mountain. The study is also based on observations with the IRAM 30m Telescope. IRAM is supported by INSU/CNRS (France), MPG (Germany) and IGN (Spain). KI\"O acknowledges funding from the Simons Collaboration on the Origins of Life Investigator award \#321183, the Alfred P. Sloan Foundation, and the David and Lucile Packard Foundation. JK's research on nearby, irradiated disks is supported by National Science Foundation grant AST-1108950 to RIT.


\end{document}